\documentclass[12pt,a4paper,fleqn]{article}
\usepackage{amsmath}
\usepackage{amsfonts}

\begin{document}

\title{\textbf{A strange recursion operator for a new integrable system of coupled
Korteweg--de~Vries equations}}
\author{Ay\c{s}e Karasu~(Kalkanl\i), Atalay Karasu, S.~Yu.~Sakovich\bigskip\\
{\footnotesize Department of Physics, Middle East Technical University,}\\
{\footnotesize 06531 Ankara, Turkey\smallskip}\\
{\footnotesize akarasu@metu.edu.tr, karasu@metu.edu.tr, saks@pisem.net}}
\date{}
\maketitle

\begin{abstract}
A recursion operator is constructed for a new integrable system of coupled
Korteweg--de~Vries equations by the method of gauge-invariant description of
zero-curvature representations. This second-order recursion operator is
characterized by unusual structure of its nonlocal part.
\end{abstract}

\section{Introduction}

Recently, the Painlev\'{e} classification of coupled Korteweg--de~Vries (KdV)
equations was made in \cite{K97}, and the following new system possessing the
Painlev\'{e} property was found there:
\begin{align}
u_{t}  &  =-\frac{\delta c_{1}}{12}u_{xxx}+\frac{3\delta c_{1}^{2}}{4d_{1}%
}v_{xxx}+c_{1}u_{x}u-\frac{3c_{1}^{2}}{d_{1}}u_{x}v-\frac{6c_{1}^{2}}{d_{1}%
}v_{x}u,\nonumber\\
v_{t}  &  =-\frac{\delta d_{1}}{12}u_{xxx}-\frac{7\delta c_{1}}{12}%
v_{xxx}+d_{1}u_{x}u-c_{1}u_{x}v-2c_{1}v_{x}u-\frac{6c_{1}^{2}}{d_{1}}v_{x}v,
\label{e19}%
\end{align}
where $\delta$, $c_{1}$ and $d_{1}$ are arbitrary nonzero constants.

More recently, the coupled KdV equations (\ref{e19}) appeared in \cite{S99},
also as a system possessing the Painlev\'{e} property, in the following form:
\begin{align}
u_{t}  &  =u_{xxx}-9v_{xxx}-12uu_{x}+18vu_{x}+18uv_{x},\nonumber\\
v_{t}  &  =u_{xxx}-5v_{xxx}-12uu_{x}+12vu_{x}+6uv_{x}+18vv_{x}, \label{eiv}%
\end{align}
which is related to (\ref{e19}) by the transformation (\ref{e19})$\mapsto
$(\ref{eiv}):
\begin{equation}
u\mapsto\delta u-\frac{3\delta}{2}v,\quad v\mapsto-\frac{\delta d_{1}}{2c_{1}%
}v,\quad t\mapsto\frac{6}{\delta c_{1}}t. \label{19iv}%
\end{equation}
It was found in \cite{S99} that the system (\ref{eiv}) admits the
zero-curvature representation (ZCR)
\begin{equation}
D_{t}X=D_{x}T-\left[  X,T\right]  , \label{zcr}%
\end{equation}
which is the compatibility condition for the overdetermined linear system
\begin{equation}
\Psi_{x}=X\Psi,\quad\Psi_{t}=T\Psi, \label{lin}%
\end{equation}
with
\begin{equation}
X=%
\begin{pmatrix}
\sigma & u-v & \sigma\left(  \frac{8}{3}u-4v\right) \\
\frac{3}{2} & -2\sigma & u-v\\
0 & \frac{3}{2} & \sigma
\end{pmatrix}
, \label{xiv}%
\end{equation}
where $\sigma$ is an essential (`spectral') parameter, $D_{t}$ and $D_{x}$
stand for the total derivatives, the square brackets denote the commutator,
$\Psi\left(  x,t\right)  $ is a three-component column, and an explicit (but
cumbersome) expression for the $3\times3$ matrix $T$ can be seen in \cite{S99}.

In the present paper, we study this new integrable system of coupled KdV
equations in its equivalent form, which is characterized by more transparent
structure of $X$ in its ZCR. Namely, making the transformation
\begin{align}
u  &  \mapsto2u-\frac{1}{3}v,\quad v\mapsto\frac{4}{3}u-\frac{1}{3}v,\quad
t\mapsto-\frac{1}{2}t,\nonumber\\
\Psi &  \mapsto S\Psi,\quad X\mapsto SXS^{-1},\quad T\mapsto STS^{-1},\quad
S=\operatorname{diag}\left(  \frac{2}{3},1,\frac{3}{2}\right)  \label{tran}%
\end{align}
in (\ref{eiv}), (\ref{lin}) and (\ref{xiv}), we obtain the system
\begin{align}
u_{t}  &  =4u_{xxx}-v_{xxx}-12uu_{x}+vu_{x}+2uv_{x},\nonumber\\
v_{t}  &  =9u_{xxx}-2v_{xxx}-12vu_{x}-6uv_{x}+4vv_{x} \label{sys}%
\end{align}
which admits the ZCR (\ref{zcr}) with
\begin{equation}
X=%
\begin{pmatrix}
\sigma & u & \sigma v\\
1 & -2\sigma & u\\
0 & 1 & \sigma
\end{pmatrix}
\label{eks}%
\end{equation}
(the explicit expression for $T$ is omitted, since not used in what follows).

Our aim is to obtain a recursion operator for the system (\ref{sys}). We
derive the recursion operator from the matrix $X$ (\ref{eks}) of the system's
ZCR. We do this, following the method of gauge-invariant description of ZCRs
of evolution equations, introduced in \cite{S95} (note also a more general
gauge-invariant formulation of ZCRs, developed in \cite{Mar}).

The paper is organized as follows. In Section~2, we find the complete class of
coupled evolution equations admitting ZCRs (\ref{zcr}) with the matrix $X$
given by (\ref{eks}). This class is determined implicitly, through a
three-term recurrence, which is explicitly solved in Section~3, giving rise to
the recursion operator sought. In Section~4, we prove that this operator
satisfies the sufficient condition of being a recursion operator in the sense
of symmetries. Section~5 contains discussion of the results.

\section{Class of systems}

Let us find the complete class of coupled evolution equations
\begin{align}
u_{t}  &  =f\left(  x,t,u,\ldots,u_{x\ldots x},v,\ldots,v_{x\ldots x}\right)
,\nonumber\\
v_{t}  &  =g\left(  x,t,u,\ldots,u_{x\ldots x},v,\ldots,v_{x\ldots x}\right)
\label{evo}%
\end{align}
which admit ZCRs (\ref{zcr}) with the matrix $X$ given by (\ref{eks}) and
without any restrictions imposed on the $3\times3$ matrix $T(x,t,u,\ldots
,u_{x\ldots x},v,\ldots,v_{x\ldots x},\sigma)$. We solve this problem
algorithmically, using the technique of cyclic bases of ZCRs, developed in
\cite{S95} (see also \cite{S02} for more examples).

Since the matrix $X$ (\ref{eks}) does not contain derivatives of $u$ and $v$,
the characteristic matrices of the ZCR are simply $C_{u}=\partial X/\partial
u$ and $C_{v}=\partial X/\partial v$:
\begin{equation}
C_{u}=%
\begin{pmatrix}
0 & 1 & 0\\
0 & 0 & 1\\
0 & 0 & 0
\end{pmatrix}
,\quad C_{v}=%
\begin{pmatrix}
0 & 0 & \sigma\\
0 & 0 & 0\\
0 & 0 & 0
\end{pmatrix}
. \label{cha}%
\end{equation}
Using the operator $\nabla_{x}$, defined as $\nabla_{x}Y=D_{x}Y-\left[
X,Y\right]  $ for any (here, $3\times3$) matrix $Y$, we compute $\nabla
_{x}C_{u}$, $\nabla_{x}C_{v}$, $\nabla_{x}^{2}C_{u}$, $\nabla_{x}^{2}C_{v}$,
etc., and find that the cyclic basis (i.e.~the maximal sequence of linearly
independent matrices $\nabla_{x}^{i}C_{u}$ and $\nabla_{x}^{j}C_{v}$,
$i,j=0,1,2,\ldots$) is
\begin{equation}
\left\{  C_{u},C_{v},\nabla_{x}C_{u},\nabla_{x}C_{v},\nabla_{x}^{2}%
C_{u},\nabla_{x}^{2}C_{v},\nabla_{x}^{3}C_{u},\nabla_{x}^{4}C_{u}\right\}  ,
\label{cyc}%
\end{equation}
and that the closure equations for this cyclic basis are
\begin{align}
\nabla_{x}^{5}C_{u}  &  =a_{1}C_{u}+a_{2}C_{v}+a_{3}\nabla_{x}C_{u}%
+a_{4}\nabla_{x}C_{v}+a_{5}\nabla_{x}^{2}C_{u}+a_{6}\nabla_{x}^{2}%
C_{v}\nonumber\\
&  +a_{7}\nabla_{x}^{3}C_{u}+a_{8}\nabla_{x}^{4}C_{u},\nonumber\\
\nabla_{x}^{3}C_{v}  &  =b_{1}C_{u}+b_{2}C_{v}+b_{3}\nabla_{x}C_{u}%
+b_{4}\nabla_{x}C_{v}+b_{5}\nabla_{x}^{2}C_{u}+b_{6}\nabla_{x}^{2}%
C_{v}\nonumber\\
&  +b_{7}\nabla_{x}^{3}C_{u}+b_{8}\nabla_{x}^{4}C_{u}, \label{clo}%
\end{align}
where
\begin{align}
a_{1}  &  =u_{xxx}-11uu_{x}+\frac{3}{2}vu_{x}-27\sigma^{2}u_{x}+6\sigma
^{2}v_{x},\nonumber\\
a_{2}  &  =-6u_{xxx}+2v_{xxx}+54uu_{x}-3vu_{x}-22uv_{x}+3vv_{x}\nonumber\\
&  +54\sigma^{2}u_{x}-18\sigma^{2}v_{x},\nonumber\\
a_{3}  &  =4u_{xx}-22u^{2}+3uv-99\sigma^{2}u+\frac{45}{2}\sigma^{2}%
v-81\sigma^{4},\nonumber\\
a_{4}  &  =-24u_{xx}+7v_{xx}+54u^{2}-36uv+\frac{9}{2}v^{2}+54\sigma
^{2}u-9\sigma^{2}v,\nonumber\\
a_{5}  &  =5u_{x},\quad a_{6}=-36u_{x}+8v_{x},\nonumber\\
a_{7}  &  =13u-\frac{3}{2}v+18\sigma^{2},\quad a_{8}=0,\nonumber\\
b_{1}  &  =\frac{1}{2}u_{x},\quad b_{2}=-u_{x}+v_{x},\quad b_{3}=u+\frac{3}%
{2}\sigma^{2},\nonumber\\
b_{4}  &  =-u+\frac{3}{2}v,\quad b_{7}=-\frac{1}{2},\quad b_{5}=b_{6}=b_{8}=0.
\label{coe}%
\end{align}
Thus, in the case of the matrix $X$ (\ref{eks}), the cyclic basis is
eight-dimensional. Therefore the singular basis (see \cite{S95} for the
definition) can be one-dimensional at most; in fact, it consists of the unit
matrix $I=\operatorname{diag}\left(  1,1,1\right)  $ and, due to $\nabla
_{x}I=0$, has no effect on the form of the represented equations (\ref{evo}).

Now, we rewrite the ZCR (\ref{zcr}) in its equivalent (characteristic) form
\begin{equation}
fC_{u}+gC_{v}=\nabla_{x}T \label{ide}%
\end{equation}
and decompose the matrix $T$ over the cyclic basis as
\begin{align}
T  &  =p_{1}C_{u}+p_{2}C_{v}+p_{3}\nabla_{x}C_{u}+p_{4}\nabla_{x}C_{v}%
+p_{5}\nabla_{x}^{2}C_{u}+p_{6}\nabla_{x}^{2}C_{v}\nonumber\\
&  +p_{7}\nabla_{x}^{3}C_{u}+p_{8}\nabla_{x}^{4}C_{u}, \label{dec}%
\end{align}
where the coefficients $p_{1},\ldots,p_{8}$ are functions of $x,t,u,\ldots
,u_{x\ldots x},v,\ldots,v_{x\ldots x}$ and $\sigma$. Taking into account the
closure equations (\ref{clo}) and the linear independence of the matrices
(\ref{cyc}), we obtain from (\ref{ide}) and (\ref{dec}) the following
relations:
\begin{align}
f  &  =D_{x}p_{1}+a_{1}p_{8}+b_{1}p_{6},\quad g=D_{x}p_{2}+a_{2}p_{8}%
+b_{2}p_{6},\nonumber\\
p_{1}  &  =-D_{x}p_{3}-a_{3}p_{8}-b_{3}p_{6},\quad p_{2}=-D_{x}p_{4}%
-a_{4}p_{8}-b_{4}p_{6},\nonumber\\
p_{3}  &  =-D_{x}p_{5}-a_{5}p_{8}-b_{5}p_{6},\quad p_{4}=-D_{x}p_{6}%
-a_{6}p_{8}-b_{6}p_{6},\nonumber\\
p_{5}  &  =-D_{x}p_{7}-a_{7}p_{8}-b_{7}p_{6},\quad p_{7}=-D_{x}p_{8}%
-a_{8}p_{8}-b_{8}p_{6}. \label{fgp}%
\end{align}

These relations (\ref{fgp}), together with (\ref{coe}), determine $p_{1}%
,p_{2},p_{3},p_{4},p_{5},p_{7},f$ and $g$ in terms of some linear differential
operators applied to the two (still undetermined) functions $p_{6}$ and
$p_{8}$. For what follows, we need no expression for $T$: it is sufficient to
know that $T$ exists. The expressions for the right-hand sides $f$ and $g$ of
the represented equations (\ref{evo}) are
\begin{equation}
h=\left(  M+\lambda L+\lambda^{2}K\right)  r, \label{rhs}%
\end{equation}
where
\begin{equation}
h=%
\begin{pmatrix}
f\\
g
\end{pmatrix}
,\quad\lambda=9\sigma^{2},\quad r=%
\begin{pmatrix}
p_{8}\\
p_{6}%
\end{pmatrix}
, \label{col}%
\end{equation}%
\begin{equation}
K=%
\begin{pmatrix}
D_{x} & 0\\
0 & 0
\end{pmatrix}
, \label{opk}%
\end{equation}%
\begin{equation}
L=%
\begin{pmatrix}
-2D_{x}^{3}+\left(  11u-\frac{5}{2}v\right)  D_{x}+\left(  8u_{x}-\frac{11}%
{6}v_{x}\right)  & -\frac{1}{6}D_{x}\\
\left(  -6u+v\right)  D_{x}-v_{x} & 0
\end{pmatrix}
, \label{opl}%
\end{equation}%
\begin{equation}
M=%
\begin{pmatrix}
M_{11} & M_{12}\\
M_{21} & M_{22}%
\end{pmatrix}
\label{opm}%
\end{equation}
with
\begin{align}
M_{11}  &  =D_{x}^{5}+\left(  -13u+\frac{3}{2}v\right)  D_{x}^{3}+\left(
-34u_{x}+\frac{9}{2}v_{x}\right)  D_{x}^{2}\nonumber\\
&  +\left(  -33u_{xx}+\frac{9}{2}v_{xx}+22u^{2}-3uv\right)  D_{x}\nonumber\\
&  +\left(  -11u_{xxx}+\frac{3}{2}v_{xxx}+33uu_{x}-\frac{3}{2}vu_{x}%
-3uv_{x}\right)  , \label{m11}%
\end{align}%
\begin{equation}
M_{12}=\frac{1}{2}D_{x}^{3}-uD_{x}-\frac{1}{2}u_{x}, \label{m12}%
\end{equation}%
\begin{align}
M_{21}  &  =\left(  -36u_{x}+8v_{x}\right)  D_{x}^{2}\nonumber\\
&  +\left(  -48u_{xx}+9v_{xx}-54u^{2}+36uv-\frac{9}{2}v^{2}\right)
D_{x}\nonumber\\
&  +\left(  -18u_{xxx}+3v_{xxx}-54uu_{x}+33vu_{x}+14uv_{x}-6vv_{x}\right)  ,
\label{m21}%
\end{align}%
\begin{equation}
M_{22}=D_{x}^{3}+\left(  u-\frac{3}{2}v\right)  D_{x}-\frac{1}{2}v_{x}.
\label{m22}%
\end{equation}

If $\sigma$ in (\ref{eks}) is a fixed constant, then (\ref{rhs}) and
(\ref{col})--(\ref{m22}) give the solution of our problem: a continual class
of coupled evolution equations, containing two arbitrary functions $p_{6}$ and
$p_{8}$, is represented by (\ref{zcr}) with (\ref{eks}). However, $\sigma$ is
a free parameter in (\ref{eks}), and we need to take into account that the
condition $\partial h/\partial\lambda=0$ must be satisfied for (\ref{rhs}).
Using the series expansion of $r$
\begin{equation}
r=r_{0}+\lambda r_{1}+\lambda^{2}r_{2}+\lambda^{3}r_{3}+\lambda^{4}%
r_{4}+\cdots\label{exp}%
\end{equation}
(if a singularity is suspected in $r$ at $\lambda=0$, one may use (\ref{exp})
after an infinitesimal shift of $\lambda$, which does not affect final
results), we obtain from (\ref{rhs}) that
\begin{equation}
h=Mr_{0} \label{mr0}%
\end{equation}
and
\begin{subequations}
\label{3tr}%
\begin{align}
0  &  =Mr_{1}+Lr_{0},\label{3tr1}\\
0  &  =Mr_{2}+Lr_{1}+Kr_{0},\label{3tr2}\\
0  &  =Mr_{3}+Lr_{2}+Kr_{1},\label{3tr3}\\
0  &  =Mr_{4}+Lr_{3}+Kr_{2},\label{3tr4}\\
&  \cdots.\nonumber
\end{align}
\end{subequations}
This gives us the following implicit solution of our problem: the system of
coupled evolution equations (\ref{evo}) admits the ZCR (\ref{zcr}) with the
matrix $X$ given by (\ref{eks}) if and only if its right-hand side $h=\left(
f,g\right)  ^{\mathrm{T}}$ is determined through (\ref{mr0}) by the set of
two-component functions $r_{0},r_{1},r_{2},\ldots$ satisfying the three-term
recurrence (\ref{3tr}), where the matrix differential operators $M$, $L$ and
$K$ are defined by (\ref{opk})--(\ref{m22}).

\section{Recursion of systems}

Let us solve the recurrence (\ref{3tr}), using formal inversion of
differential operators. We say that the operator $P^{-1}$ is formally inverse
to a linear differential operator $P$ if $P^{-1}P=PP^{-1}=I$, where $I$ is the
unit operator. In other words, $P^{-1}a$ denotes any function $b$ of local
variables, such that $a=Pb$, if $b$ exists. We have to remind that, in the
present case, differential operators contain total (not partial) derivatives
$D_{x}$ and act on differential functions (in Olver's sense \cite{Olv},
i.e.~functions of independent variables, dependent variables and finite-order
derivatives of dependent variables). For this reason, $D_{x}^{-1}0=\phi\left(
t\right)  $ with any function $\phi$, whereas $\left(  D_{x}^{2}+u\right)
^{-1}0=0$ and $\left(  3D_{x}^{3}-4vD_{x}-2v_{x}\right)  ^{-1}v_{x}=-\frac
{1}{2}$, to give some examples.

For the operators $K$, $L$ and $M$, defined by
(\ref{opk})--(\ref{m22}), we notice that the inverse operators
$M^{-1}$ and $L^{-1}$ exist, but $K^{-1}$ does not exist, and
therefore we cannot use inversion of $K$ in what follows. We
obtain
\begin{equation}
r_{0}=-L^{-1}Mr_{1} \label{r01}%
\end{equation}
from (\ref{3tr1}), and
\begin{equation}
r_{1}=\left(  KL^{-1}M-L\right)  ^{-1}Mr_{2} \label{r12}%
\end{equation}
from (\ref{3tr2}), and notice that the form of (\ref{r12}) is different from
the form of (\ref{r01}), because $KL^{-1}M\neq0$ for the operators defined by
(\ref{opk})--(\ref{m22}). Then we obtain
\begin{equation}
r_{2}=\left(  -K\left(  KL^{-1}M-L\right)  ^{-1}M-L\right)  ^{-1}Mr_{3}
\label{r23}%
\end{equation}
from (\ref{3tr3}), and find that, fortunately, it is possible to rewrite
(\ref{r23}) as
\begin{equation}
r_{2}=\left(  KL^{-1}M-L\right)  ^{-1}Mr_{3} \label{f23}%
\end{equation}
owing to the condition
\begin{equation}
KL^{-1}K=0 \label{klk}%
\end{equation}
satisfied by $K$ (\ref{opk}) and $L$ (\ref{opl}). We see that the form of
(\ref{f23}) coincides with the form of (\ref{r12}). The same is true, also
owing to (\ref{klk}), for the relation between $r_{3}$ and $r_{4}$, which
follows from (\ref{3tr4}); and, in general,
\begin{equation}
r_{i}=\left(  KL^{-1}M-L\right)  ^{-1}Mr_{i+1},\quad i=1,2,3,\ldots.
\label{ii1}%
\end{equation}
We have to emphasize that the recursion (\ref{ii1}) is caused by the specific
property (\ref{klk}) of the operators $K$ (\ref{opk}) and $L$ (\ref{opl}), and
that it would be impossible to solve the three-term recurrence (\ref{3tr}) for
generic operators.

For convenience in what follows, we rewrite (\ref{ii1}) as
\begin{equation}
r_{i}=M^{-1}LM^{-1}RML^{-1}Mr_{i+1},\quad i=1,2,3,\ldots, \label{rew}%
\end{equation}
where
\begin{equation}
R=MN^{-1} \label{der}%
\end{equation}
with
\begin{equation}
N=ML^{-1}K-L. \label{den}%
\end{equation}
The explicit expressions for $N$ and $R$ are
\begin{equation}
N=%
\begin{pmatrix}
-D_{x}^{3}+\left(  -5u+\frac{5}{2}v\right)  D_{x}+\left(  -5u_{x}+\frac{11}%
{6}v_{x}\right)  & \frac{1}{6}D_{x}\\
-6D_{x}^{3}+8vD_{x}+4v_{x} & 0
\end{pmatrix}
\label{opn}%
\end{equation}
and
\begin{equation}
R=%
\begin{pmatrix}
R_{11} & R_{12}\\
R_{21} & R_{22}%
\end{pmatrix}
\label{mar}%
\end{equation}
with
\begin{equation}
R_{11}=3D_{x}^{2}-6u-3u_{x}D_{x}^{-1}, \label{o11}%
\end{equation}%
\begin{align}
R_{12}  &  =\left[  -2D_{x}^{5}+\left(  2u+3v\right)  D_{x}^{3}+\left(
-4u_{x}+8v_{x}\right)  D_{x}^{2}\right. \nonumber\\
&  +\left(  -6u_{xx}+7v_{xx}+4u^{2}-6uv\right)  D_{x}\nonumber\\
&  +\left(  -2u_{xxx}+2v_{xxx}+6uu_{x}-3vu_{x}-4uv_{x}\right) \nonumber\\
&  \left.  +u_{x}D_{x}^{-1}v_{x}\right]  \left(  3D_{x}^{3}-4vD_{x}%
-2v_{x}\right)  ^{-1}, \label{o12}%
\end{align}%
\begin{equation}
R_{21}=6D_{x}^{2}+\left(  6u-9v\right)  -3v_{x}D_{x}^{-1}, \label{o21}%
\end{equation}%
\begin{align}
R_{22}  &  =\left[  -3D_{x}^{5}+\left(  -18u+12v\right)  D_{x}^{3}+\left(
-27u_{x}+18v_{x}\right)  D_{x}^{2}\right. \nonumber\\
&  +\left(  -21u_{xx}+14v_{xx}+12u^{2}+12uv-9v^{2}\right)  D_{x}\nonumber\\
&  +\left(  -6u_{xxx}+4v_{xxx}+12uu_{x}+6vu_{x}+6uv_{x}-9vv_{x}\right)
\nonumber\\
&  \left.  +v_{x}D_{x}^{-1}v_{x}\right]  \left(  3D_{x}^{3}-4vD_{x}%
-2v_{x}\right)  ^{-1}. \label{o22}%
\end{align}

Now, using (\ref{mr0}), (\ref{r01}) and (\ref{rew}), we obtain the following
expression for $h$:
\begin{equation}
h=-R^{n}ML^{-1}Mr_{n+1},\quad n=0,1,2,\ldots. \label{hn1}%
\end{equation}
Since the leading-order terms of the operator $R$ (\ref{mar})--(\ref{o22}) are
given by
\begin{equation}
R=%
\begin{pmatrix}
3D_{x}^{2}+\cdots & -\frac{2}{3}D_{x}^{2}+\cdots\\
6D_{x}^{2}+\cdots & -D_{x}^{2}+\cdots
\end{pmatrix}
\label{lot}%
\end{equation}
and the orders of derivatives of $u$ and $v$ in $h$ must be finite, we can
choose a sufficiently large $n$ in (\ref{hn1}), so that the orders of
derivatives of $u$ and $v$ in $ML^{-1}Mr_{n+1}$ are less than 2 for this $n$.
In this case, we conclude, taking into account the form of $M$ (\ref{opm}%
)--(\ref{m22}) and $L$ (\ref{opl}), that $L^{-1}Mr_{n+1}$ must be a function
of $x$ and $t$ only, that the orders of derivatives of $u$ and $v$ in
$Mr_{n+1}$ must be less than 2, and that $r_{n+1}=\left(  0,0\right)
^{\mathrm{T}}$. Then we find that
\begin{equation}
L^{-1}%
\begin{pmatrix}
0\\
0
\end{pmatrix}
=%
\begin{pmatrix}
0\\
2\phi\left(  t\right)
\end{pmatrix}
, \label{inl}%
\end{equation}
where the function $\phi\left(  t\right)  $ is arbitrary, and the factor 2 is
taken for convenience. Finally, (\ref{hn1}) gives us
\begin{equation}
h=\phi\left(  t\right)  R^{n}%
\begin{pmatrix}
u_{x}\\
v_{x}%
\end{pmatrix}
,\quad\forall\phi\left(  t\right)  ,\quad n=0,1,2,\ldots, \label{hnx}%
\end{equation}
or, equivalently,
\begin{equation}
h=R^{n}%
\begin{pmatrix}
0\\
0
\end{pmatrix}
,\quad n=1,2,3,\ldots, \label{hie}%
\end{equation}
because $R\left(  0,0\right)  ^{\mathrm{T}}=\left(  \phi\left(  t\right)
u_{x},\phi\left(  t\right)  v_{x}\right)  ^{\mathrm{T}}$ with any $\phi\left(
t\right)  $.

Thus, the problem, formulated in the beginning of Section~2, is solved
explicitly: the relation (\ref{hie}) determines the right-hand sides
$h=\left(  f,g\right)  ^{\mathrm{T}}$ of all systems (\ref{evo}) which admit
the linear problem (\ref{lin}) with the matrix $X$ (\ref{eks}). The
constructed operator $R$ (\ref{mar})--(\ref{o22}) turns out to be a recursion
operator in the Lax sense: it generates a hierarchy of integrable systems
which all possess Lax pairs with the predetermined spatial part. The system
(\ref{sys}) corresponds to $n=2$ in (\ref{hie}), and the next member of the
hierarchy ($n=3$) is the fifth-order system
\begin{align}
u_{t}  &  =6u_{xxxxx}-\frac{5}{3}v_{xxxxx}-60uu_{xxx}+\frac{40}{3}%
vu_{xxx}+\frac{50}{3}uv_{xxx}\nonumber\\
&  -\frac{10}{3}vv_{xxx}-150u_{x}u_{xx}+40v_{x}u_{xx}+\frac{125}{3}u_{x}%
v_{xx}-10v_{x}v_{xx}\nonumber\\
&  +120u^{2}u_{x}-40uvu_{x}+\frac{5}{3}v^{2}u_{x}-\frac{80}{3}u^{2}v_{x}%
+\frac{20}{3}uvv_{x},\nonumber\\
v_{t}  &  =15u_{xxxxx}-4v_{xxxxx}-120uu_{xxx}+10vu_{xxx}+30uv_{xxx}\nonumber\\
&  -\frac{5}{3}vv_{xxx}-360u_{x}u_{xx}+80v_{x}u_{xx}+90u_{x}v_{xx}-\frac
{55}{3}v_{x}v_{xx}\nonumber\\
&  +160uvu_{x}-40v^{2}u_{x}+40u^{2}v_{x}-60uvv_{x}+\frac{35}{3}v^{2}v_{x}.
\label{5th}%
\end{align}

\section{Recursion of symmetries}

The operator $R$, defined by (\ref{mar})--(\ref{o22}), is also a recursion
operator with respect to symmetries of the coupled KdV equations (\ref{sys}).
In order to see this, we have to prove that $R$ satisfies the following
condition \cite{Olv}:
\begin{equation}
D_{t}R=\left[  H,R\right]  , \label{sym}%
\end{equation}
where $H$ is the Fr\'{e}chet derivative of the right-hand side of the system
(\ref{sys}),
\begin{equation}
H=%
\begin{pmatrix}
H_{11} & H_{12}\\
H_{21} & H_{22}%
\end{pmatrix}
\label{fre}%
\end{equation}
with
\begin{align}
H_{11}  &  =4D_{x}^{3}+\left(  -12u+v\right)  D_{x}+\left(  -12u_{x}%
+2v_{x}\right)  ,\nonumber\\
H_{12}  &  =-D_{x}^{3}+2uD_{x}+u_{x},\nonumber\\
H_{21}  &  =9D_{x}^{3}-12vD_{x}-6v_{x},\nonumber\\
H_{22}  &  =-2D_{x}^{3}+\left(  -6u+4v\right)  D_{x}+\left(  -12u_{x}%
+4v_{x}\right)  . \label{hij}%
\end{align}
It is convenient to study the condition (\ref{sym}) in its equivalent form:
\begin{equation}
HM-D_{t}M=R\left(  HN-D_{t}N\right)  , \label{hmn}%
\end{equation}
which follows from (\ref{sym}) through the definition of $R$ (\ref{der}) and
the identity $D_{t}N^{-1}=-N^{-1}\left(  D_{t}N\right)  N^{-1}$. Direct
computations show that the condition (\ref{hmn}) is satisfied identically by
$H$ (\ref{fre})--(\ref{hij}), $M$ (\ref{opm})--(\ref{m22}), $R$ (\ref{mar}%
)--(\ref{o22}), $N$ (\ref{opn}) and the expressions (\ref{sys}) for $u_{t}$
and $v_{t}$. Since the condition (\ref{sym}) is satisfied, the system of
coupled KdV equations (\ref{sys}) possesses infinitely many symmetries: the
recursion operator $R$ generates a `new' symmetry (but not necessarily a local
one) from any `old' symmetry.

\section{Conclusion}

The recursion operator $R$ (\ref{mar})--(\ref{o22}), constructed
in this paper for the new integrable system of coupled KdV
equations (\ref{sys}), is characterized by unusual structure of
its nonlocal part, which contains not only the conventional
operator $D_{x}^{-1}$ but also the strange nonlocal operator
$\left(  3D_{x}^{3}-4vD_{x}-2v_{x}\right)  ^{-1}$. Such a
phenomenon has not been observed in the literature as yet. This
operator $\left( 3D_{x}^{3}-4vD_{x}-2v_{x}\right)  ^{-1}$ cannot
be expressed in terms of a finite number of operators
$D_{x}^{-1}$, if one uses only local variables, as we did
throughout the paper. However, if we introduce the nonlocal
variable $w:\ v=3w_{xx}/w$, we can represent the obtained
recursion operator in a more conventional form, using the factorization
$\left(  3D_{x}^{3}-4vD_{x}%
-2v_{x}\right)  ^{-1}=\frac{1}{3}w^{2}D_{x}^{-1}w^{-2}D_{x}^{-1}w^{-2}%
D_{x}^{-1}w^{2}$.

We found the recursion operator of the system (\ref{sys}) by the technique of
cyclic bases of ZCRs. It is interesting if this recursion operator can be
obtained by different methods as well, e.g.\ by those developed in
\cite{GKS,KK,MS}.

We have seen that the relation (\ref{hie}) generates local expressions for $h$
at $n=1,2,3$. It is easy to check that the seventh-order system of this
hierarchy ($n=4$) is also a local expression. Nevertheless, we have not proven
as yet that this is the case for any $n$.

Though our method produces the recursion operator in the quotient form
$MN^{-1}$ (\ref{der}), the operators $M$ (\ref{opm})--(\ref{m22}) and $N$
(\ref{opn}) are not Hamiltonian operators. It is still unknown if the obtained
recursion operator is hereditary and if the system (\ref{sys}) admits a
bi-Hamiltonian structure.

\section*{Acknowledgments}

The authors are grateful to the Scientific and Technical Research Council of
Turkey (T\"{U}B\.{I}TAK) for support. One of the authors, S.~Yu.~S., is also
grateful to the Middle East Technical University (ODT\"{U}) for hospitality.

\end{document}